\def\al{\mbox{$^{26}$\hspace{-0.2em}Al}}            
\def\na{\mbox{$^{22}$\hspace{-0.1em}Na}}            
\def\MeV{\mbox{Me\hspace{-0.1em}V}}                 
\def\keV{\mbox{ke\hspace{-0.1em}V}}                 
\def\Msol{\hbox{M$_{\odot}$}}                       
\def\deg{\hbox{$^\circ$}}                           
\def\funit{\mbox{photons cm$^{-2}$ s$^{-1}$}}       
 
\def\py{\mbox{yr$^{-1}$}}
\def\gray{\mbox{$\gamma$-ray}}                      
 
\documentstyle[epsfig]{mn}

\title[Galactic 1.275~\MeV\ emission from ONe novae and its detectability by INTEGRAL/SPI]
  {Galactic 1.275~\MeV\ emission from ONe novae and its detectability by INTEGRAL/SPI} 

\author[P.~Jean et al.]
{P.~Jean,$^{1,2}$ M.~Hernanz,$^2$ J.~G\'omez-Gomar,$^2$ J.~Jos\'e,$^{2,3}$\\
	$^1$Centre d'Etude Spatiale des Rayonnements, CNRS/UPS, 9 avenue du 
	colonel Roche, 31028 Toulouse, FRANCE\\
	$^2$Institut d'Estudis Espacials de Catalunya (IEEC), CSIC, Gran 
	Capit\`a 2-4, E-08034 Barcelona, 
	SPAIN\\
	$^3$Departament de F\'{\i}sica i Enginyeria Nuclear (UPC), Av.
	V\'{\i}ctor Balaguer, s/n, E-08800 Vilanova i la Geltr\'u (Barcelona), 
	SPAIN}

\date{Accepted 2000 March 29. Received 2000 March 7}
\pagerange{\pageref{firstpage}--\pageref{lastpage}}
\pubyear{2000}
 
\def\LaTeX{L\kern-.36em\raise.3ex\hbox{a}\kern-.15em
    T\kern-.1667em\lower.7ex\hbox{E}\kern-.125emX}

\begin{document}

\label{firstpage}

\maketitle
            

\begin{abstract}

Models of Galactic 1.275 \MeV\ emission produced by the decay of 
the radionuclide \na\ have been computed. Several frequency-spatial 
distributions of novae have been investigated using recent results of nova 
rates and spatial distributions of novae in our Galaxy. These models allow us 
to 
estimate the lower-limit of the \na\ mass ejected per ONe nova detectable 
with the future spectrometer (SPI) of the INTEGRAL observatory as a 
function of the frequency-spatial distribution of ONe novae in the Galaxy. 
Calculations using recent estimations of the expected \na\ mass ejected per 
ONe nova show that the detection of the Galactic emission 
of 1.275 \MeV\ photons will be difficult with the future spectrometer of 
the INTEGRAL observatory, whereas the cumulative emission around the Galactic 
center has some chances to be detected during the deep survey of the central 
radian of the Galaxy.

\end{abstract} 
 

\section{\label{s1}Introduction} 

Observations with several instruments have provided upper-limits to the 
1.275~\MeV\ flux, related to \na\ decay from the Galaxy or from individual 
novae, (see Clayton \& Hoyle \shortcite{CH74} for the original idea to look for \na\ lines in novae). Observations with the HEAO3 \gray\ spectrometer provided an upper limit of 4 $\times$ 10$^{-4}$ \funit\ on the Galactic accumulated flux at 1.275~\MeV\ (Higdon \& Fowler \shortcite{HF87}, hereafter 
HF87). With the \gray\ spectrometer of the Solar Maximum Mission, 
Leising et~al. \shortcite{Lei88} found a 99\%\ confidence limit of 1.2 $\times$ 10$^{-4}$ \funit\ 
on a steady 1.275~\MeV\ flux from the Galactic center direction. Iyudin et al. \shortcite{Iyu95}, using observations of single novae, with the  
COMPTEL instrument on board the Compton Gamma-Ray Observatory,  
estimated a 2$\sigma$ 
upper-limit of 3 $\times$ 10$^{-5}$ \funit\ for any neon-type nova in the 
Galactic disk, which has been translated into an upper-limit of the ejected 
\na\ mass of 3.7 10$^{-8}$ \Msol.


The next generation of \gray\ spectrometers will have the sensitivity 
required for the detection of the 1.275~\MeV\  emission from classical ONe 
novae. In particular, SPI, the future spectrometer of the INTEGRAL observatory will be 
operational in 2001. With its high energy resolution 
($\Delta$E/E $\approx$0.2\%), the spectrometer is designed for the detection of 
astrophysical \gray\ lines in the 20~\keV\ - 8~\MeV\ energy range. The 
detection plane is made of 19 germanium (Ge) detectors and is surrounded by 
an active shield made with bismuth germanate (BGO) scintillators, in order to 
reduce the instrumental background. SPI will be able to perform images 
with an angular resolution of $\approx$3\deg\ by using a coded aperture system 
made of tungsten elements. The mean fully coded field-of-view is 
$\approx$16\deg\  
(for a more detailed description of the spectrometer see Mandrou et al. \shortcite{Man97} and 
Teegarden et al. \shortcite{Tee97}). Using calculations of \na\ yields in ONe novae by Jos\'e \& Hernanz \shortcite{JH98}
and instrumental background predictions, Hernanz et al. \shortcite{Her97} and Go\'mez-Gomar et al. \shortcite{Gom98}
estimated that the 1.275~\MeV\ line from an individual ONe nova could be
detected by SPI if its distance 
is less than $\approx$ 0.5~kpc ({\it standard} CO novae produce \na\ in much 
smaller amounts than ONe novae). With a new analysis of the nuclear reaction rates involved in \na\ synthesis \cite{JCH99} a bit larger distance is obtained (around 1.5 kpc, Hernanz et al. \shortcite{Her99}).

The total Galactic flux at 1.275~\MeV\ depends not only on the amount of \na\ 
ejected per outburst but also on the distribution and the rate of Galactic ONe novae. However, the Galactic nova rate is poorly known,
independently of the underlying white dwarf composition, because the
interstellar extinction prevents us from directly observing more than a 
small fraction of the novae that explode each year. Several methods have been 
used to estimate the nova occurrence rate. Estimations based on 
extrapolations of Galactic nova observations suggest a rate in the range 
50-100 \py (see Liller \& Mayer \shortcite{LM87} and references therein). 
Observations of novae in other galaxies have revealed a 
correlation between the nova rate and the infrared luminosity of the 
parent galaxy. However, the occurrence rates of Galactic novae computed by 
scaling the nova rate measured in external galaxies are lower than 
$\approx$50 \py. With this method, Ciardullo et al. \shortcite{Cia90} predicted a nova rate in the range 11-45~\py, and Della Valle \& Livio \shortcite{DVL94} a rate $\approx$20~\py. This is significantly lower than estimations based on 
Galactic nova observations. Recently, Shafter \shortcite{Sha97} reconciled this 
difference by recomputing the nova rate with the Galactic nova data. He 
extrapolated the global nova rate from the observed one, accounting 
for surface brightnesses of the bulge and the disk components and 
correction factors taking care of any observational incompleteness. With 
this method, Shafter \shortcite{Sha97} estimated the nova rate to be 35$\pm$11 \py. Hatano et al. \shortcite{Hat97} found a similar value (41$\pm$20 \py) using a Monte-Carlo technique with a simple model for the distribution of dust and classical novae in the Galaxy. 
As only ONe novae are important producers of \na, the Galactic ONe nova rate 
is the relevant one for this work. 
The Galactic ONe nova rate is obtained by multiplying the total nova rate 
by the fraction of novae that results from thermonuclear runaways on ONe white dwarfs. Williams et al. \shortcite{Wil85} deduced from observations of abundances in nova ejecta a proportion of ONe novae from 20 to 40 percent. With a detailed model of white dwarf mass distribution in cataclysmic binaries, 
Ritter et al. \shortcite{Rit91} estimated that 25-57 percent of observed nova 
outbursts should occur on ONe white dwarfs. Livio \& Truran \shortcite{LT94} reestimated the frequency of occurrence of ONe novae, in light of observations of abundances in nova ejecta. They concluded that, of the 18 classical novae for which detailed abundance analyses were available, only two or three had a large 
amount of neon and were ONe novae, whereas three other novae showed a modest enrichment in neon, casting doubt on the type of the underlying white dwarf. Under these considerations, they estimated a fraction of Galactic ONe novae between 11 and 33 percent.

In addition to the \na\ mass ejected per outburst and the ONe novae rate, the distribution of ONe novae in the Galaxy is an important parameter that tunes the total 1.275~\MeV\ flux arriving on Earth. Some \na\ mass upper-limits, resulting from 1.275~\MeV\ total Galactic flux upper-limit measurements, are underestimated since they have been computed assuming that all the ONe novae are embedded in the Galactic Center. Indeed, a larger \na\ mass can be ejected by ONe novae leading to a 1.275~\MeV\ total Galactic flux below the measured upper-limit if the ONe nova distribution is more diluted in the Galaxy.

In this paper, we model the Galactic cumulative emission at 1.275~\MeV\ as a function 
of the ONe novae rate and the mean \na\ yield per outburst, 
for several spatial distributions of novae in the Galaxy. The 
lower-limit for the mean mass yield of \na\ ejected per nova, 
in order to detect the 1.275~\MeV\ cumulative emission, 
is predicted for the future observations of SPI, using the estimation of the 
sensitivity of the future spectrometer developed by Jean \shortcite{Jea96} and 
Jean et al. \shortcite{Jea97}. 
The method for the simulation of the Galactic 1.275~\MeV\  emission and its 
results are described 
in section 2, whereas the method of observation with SPI and its 
application to the previous simulations are 
presented in section 3. Discussion and conclusions follow (section 
4).

\section{Simulation of the Galactic 1.275~\MeV\ emission}

\subsection{Method}


The 1.275~\MeV\ emission from Galactic ONe novae is calculated with a
Monte-Carlo simulation. A simulation of a set of ONe novae that is
representative of what could be the Galactic novae distribution at a given
time, i.e. a given novae frequency-spatial distribution, is made (this 
set of novae is further called a `Galaxy-test'). 
The two nova populations (bulge and disk) 
have been considered. The population origin (bulge or disk), the
position in galacto-centric coordinates (depending on the chosen
population) and the age of ONe novae are chosen randomly according to
the appropriate distributions. The local-galactic coordinates
($l,b$) of each nova as well as its distance to the Sun are
calculated assuming a Sun to Galactic center (hereafter GC) distance of 8~kpc.
The 1.275~\MeV\  flux ($F_{22}$ in \funit) has been estimated using the
relationship 

\begin{equation}
 F_{22} \; = \; 3.17 \: 10^{5} \; \frac{M_{22} \; 
e^{-\frac{t}{\tau_{22}}}}{A_{22} \; \tau_{22} \; d^{2}} 
\label{eq:fl}
\end{equation}

\noindent
where $M_{22}$ is the \na\ mass ejected per nova (in solar masses),
$t$ the age of the nova (in years), $d$ its distance to the Sun (in kpc).
$A_{22}$ and $\tau_{22}$ are the mass number and the
lifetime of \na\ (3.75~yr), respectively. Limitation of the flux due to 
opacity of the ejecta has not been taken into account, since it is significant 
only 
in the first week after the explosion \cite{Gom98}, which is a short time 
as compared to the \na\ lifetime. The number of simulated ONe novae depends on
the global (CO and ONe) nova frequency ($f_{n}$), the total period
during which ejected \na\ is an effective emitter ($N_{\tau} \,
\tau_{22}$), the proportion
of novae in the bulge ($p_{b}$) and the proportion of ONe novae in
the bulge (with respect to all novae in the bulge, $q_{b}$) and in the disk 
(with respect to all novae in the disk, $q_{d}$). 
The number of simulated ONe novae is calculated as

\begin{equation}
 N_{n} \; = \; f_{n} \; N_{\tau} \; \tau_{22} \; \left[ p_{b} \; q_{b}
 \; + \; \left( 1 \, - \, p_{b} \right) \; q_{d} \right] 
\label{eq:nn}
\end{equation}

\noindent
It has been assumed that the emissivity of novae older than 5$\tau_{22}$
has a negligible contribution to the diffuse Galactic 1.275~\MeV\ emission; 
therefore, $N_{\tau}$=5. The proportion of ONe novae in the bulge (with respect to all ONe novae) is given by

\begin{equation}
 p_{ONe}^{bulge} \; = \; \frac{p_{b} \; q_{b}}{p_{b} \; q_{b} \; + \;
 (1 \: - \: p_{b}) \; q_{d}}  
\label{eq:pb} 
\end{equation}

The simulation of a Galaxy-test provides a set of 1.275~\MeV\  line fluxes
$F_{i}$ for each nova $i$ localized at $l_{i}$, $b_{i}$. Distribution
of the intensity and of the number of novae has been calculated with
3\deg\ by 3\deg\ size pixels, which correspond to the angular resolution
of SPI.

\subsection{\label{s22} Models}
\subsubsection{\label{s221} Distribution of novae in the Galaxy}


For the purpose of this work, a Galaxy model with a disk and a spheroid is 
convenient to simulate the distribution of Galactic nova events. Several 
laws for the spatial distribution in the disk and in the `spheroid' 
(representative of the bulge) 
have been proposed in the literature. We have selected four 
models that differ significantly from each other. The first and older of them 
is described in Higdon \& Fowler \shortcite{HF87}. The radial scalelength of the disk is derived from the starlight suface brightness distribution \cite{dVP78} and the height dependence is from Patterson \shortcite{Pat84}. HF87 use the analytic dependence for the spheroid light distribution from Bahcall et al. \shortcite{BSS82} and Young \shortcite{You76}, 
also called the `R$^{1/4}$' law (R is the distance to the GC), 
originally formulated by de Vaucouleurs \shortcite{dVa48}. By scaling the nova rate in the bulge 
of M31 to the bulge of our Galaxy, HF87 estimate a proportion of novae in the 
spheroid of 0.348. 
The second model has been used by Hatano et al. \shortcite{Hat97} to estimate the spatial 
distribution and the occurrence rate of Galactic classical novae. It is 
based on a model of the distribution of type Ia supernovae \cite{DJ94}. 
The proportion of novae that 
occur in the bulge is set to 0.111 on the basis of an estimate of the 
bulge to total galaxy mass ratio.
The third model is derived from the Galactic survey of the Spacelab 
InfraRed Telescope (IRT) that provides a reliable tracer of the 
distribution of G and K giant stars (see Kent, Dame \& Fazio \shortcite{KDF91} and Kent \shortcite{Kent92}). 
It has also been used by Prantzos \& Diehl \shortcite{PD96} to estimate the contribution of old 
population stars (novae and AGB) to the Galactic \al\ emission at 1.8 
\MeV\ measured by COMPTEL. Using the total infrared luminosity of the bulge 
and the disk, the derived proportion of novae occurring in the bulge is 0.179. 
The last model is taken from Van der Kruit \shortcite{VdK90}. It has been used by Shafter \shortcite{Sha97} to estimate the nova rate in our Galaxy. This author assumes that the nova distribution follows the brightness profile of our Galaxy. Under this assumption, the proportion of bulge novae is 0.105. 

\begin{figure}
 \psfig{file=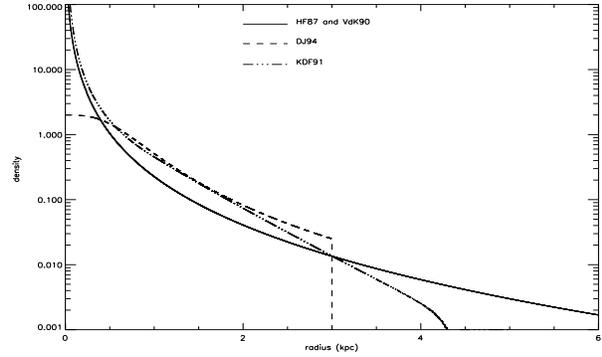,height=4.9cm,width=8.0cm}
 \caption{Radial distributions of novae in the Galactic bulge for four models (in kpc$^{-3}$). HF87: model used by Higdon and Fowler (1987) derived 
from the spheroid light distribution. This last model has also been adopted by 
Van der Kruit, 1990 (VdK90). KDF91: model from Kent, Dame and Fazio
(1994) corresponding to the distribution of G and K giant stars
provided by the InfraRed Telescope observations. DJ94: model from Dawson
and Johnson (1994) derived from the distribution of type Ia supernovae.
This last model has been truncated at 3~kpc (see table \ref{tab:model} for 
details).}
 \label{fig:sphe}
\end{figure}

\begin{figure}
 \psfig{file=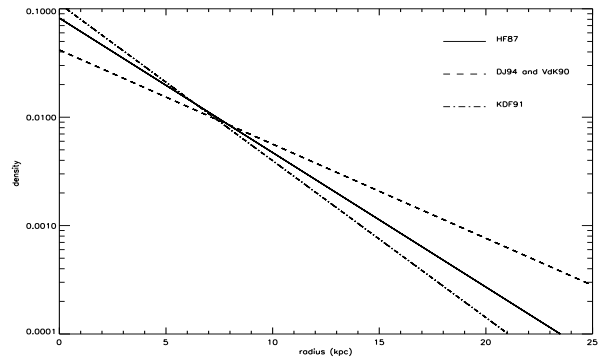,height=4.9cm,width=8.0cm} 
 \caption{Radial distributions (cylindrical coordinate) of novae in the Galactic disk (in kpc$^{-2}$) for the four models discussed in figure \ref{fig:sphe} (see text for details).}
 \label{fig:diskr}
\end{figure}

\begin{figure}
 \psfig{file=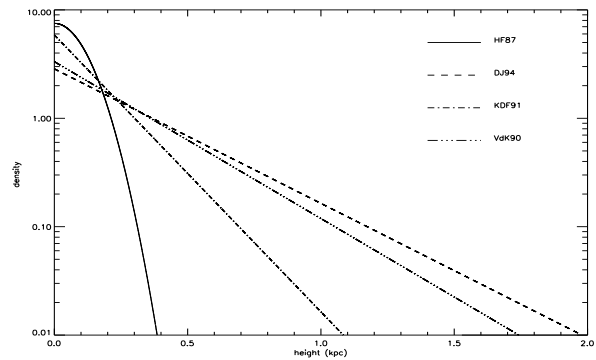,height=4.9cm,width=8.0cm} 
 \caption{Height above the Galactic plane distributions (cylindrical coordinate) of novae in the Galactic disk (in kpc$^{-1}$) for the four models discussed in figure \ref{fig:sphe} (see text for details).}
 \label{fig:diskz}
\end{figure}

The adopted distributions are shown in Figures \ref{fig:sphe}, \ref{fig:diskr} 
and \ref{fig:diskz}, which represent the probability to find a nova in the 
Galactic bulge at spherical radius R in a volume element (kpc$^{-3}$, Figure 
\ref{fig:sphe}), and the probabilities to find a nova in the Galactic disk at 
cylindrical radius $\rho$ in an area element (kpc$^{-2}$, Figure 
\ref{fig:diskr}) and at height z above the Galactic plane in a line element 
(kpc$^{-1}$, Figure \ref{fig:diskz}). Table \ref{tab:model} summarizes the characteristics of the adopted models.

\begin{table*}

\begin{minipage}{12cm}

\caption{Nova spatial distributions used for the simulation of
the Galactic 1.275 \MeV\ emission. $R$ is the
distance to the GC, z is the distance perpendicular to
the Galactic plane and $\rho$ is the galactocentric planar distance. The
distance from the GC to the Sun is $R_{\odot}$=8~kpc, $n_s$ and $n_d$ are
the normalization factors for the spheroid and the disk respectively (in
kpc$^{-3}$). The proportions of novae in the spheroid, $p_b$, are 0.348 
(HF87), 0.105 (VdK90), 0.179 (KDF91) and 0.111 (DJ94)}

\label{tab:model}


\mbox{Model HF87: $\rho_h$=3.5~kpc and $z_o$=0.106~kpc.}
\begin{tabular}{lclr}

Disk $n(z,\rho)$ & = & $n_{d}
e^{-\frac{1}{2}\frac{z^{2}}{z_{o}^{2}} - \frac{\rho}{\rho_{h}}}$
&                  \\
Bulge $n(R)$ & = & $n_{s} 1.25 \left(
\frac{R}{R_{\odot}} \right)
^{-\frac{6}{8}}e^{-10.093(\frac{R}{R_{\odot}})^{\frac{1}{4}}+10.093}$ &
$R \leq R_{\odot}$ \\
 & = & $n_{s} \left( \frac{R}{R_{\odot}}\right) ^{-\frac{7}{8}} \left[
 1-\frac{0.0867}{(\frac{R}{R_{\odot}})^{\frac{1}{4}}} \right]
 e^{-10.093(\frac{R}{R_{\odot}})^{\frac{1}{4}}+10.093}$ & $R \geq
 R_{\odot}$

\end{tabular}



\mbox{Model VdK90: $\rho_h$=5.0~kpc and $z_h$=0.30~kpc.}
\begin{tabular}{lclr}

Disk $n(z,\rho)$ & = & $n_{d} e^{-\frac{|z|}{z_h} - \frac{\rho}{\rho_{h}}}$
&                  \\
Bulge $n(R)$ & = & $n_{s} 1.25 \left(
\frac{R}{R_{\odot}} \right) ^{-\frac{6}{8}}e^{-10.093(\frac{R}{R_{\odot}})^{\frac{1}{4}}+10.093}$ &
$R \leq R_{\odot}$ \\
  & = & $n_{s} \left( \frac{R}{R_{\odot}}\right) ^{-\frac{7}{8}} \left[
 1-\frac{0.0867}{(\frac{R}{R_{\odot}})^{\frac{1}{4}}} \right]
 e^{-10.093(\frac{R}{R_{\odot}})^{\frac{1}{4}}+10.093}$ & $R \geq
 R_{\odot}$

\end{tabular}



\mbox{Model KDF91: $\rho_h$=3.0~kpc and $z_h$=0.170~kpc. K$_0$ is the
modified Bessel function.}
\begin{tabular}{lclr}

Disk $n(z,\rho)$ & = & $n_{d} e^{-\frac{|z|}{z_h} - \frac{\rho}{\rho_{h}}}$ &                  \\

Bulge $n(R)$ & = & $n_{s} \; 1.04 \times 10^6 \left( \frac{R}{0.482}
\right)^{-1.85}$ & $R \leq 0.938 kpc$ \\

	& = & $n_{s} \; 3.53 \; K_0 \left( \frac{R}{0.667} \right)$ & $R \geq
0.938 kpc$ \\
  & = & 0 & $R \geq 5 kpc$

\end{tabular}



\mbox{Model DJ94: $\rho_h$=5.0~kpc and $z_h$=0.35~kpc.}
\begin{tabular}{lclr}

Disk $n(z,\rho)$ & = & $n_{d} e^{-\frac{|z|}{z_h} -
\frac{\rho}{\rho_{h}}}$ &  \\

Bulge $n(R)$ & = & $\frac{n_{s}}{R^3+0.343}$ & $R \leq 3 kpc$ \\

	& = & 0 & $R \geq 3 kpc$

\end{tabular}


\end{minipage}

\end{table*}



\subsubsection{\label{s222} Galactic ONe nova rate}

As presented in section 1, methods used to estimate the occurrence rate of novae are based on extragalactic and Galactic nova observations. 
The first method can underestimate the Galactic nova rate since the nova 
count can be biased by the extinction of the Galactic disk.
On the other hand, estimations obtained with Galactic nova observations 
depend on the Galaxy model used for the extrapolation. The last estimations 
of classical novae rates are those by Shafter \shortcite{Sha97} (35$\pm$11 \py) and Hatano et al. \shortcite{Hat97} (40$\pm$20 \py). 
Therefore, for the calculations presented in section 2.3, we have adopted
frequencies of Galactic novae ranging from 20~\py\ to 60~\py\ and a proportion 
of ONe novae from 10~\%\ to 30~\%, in agreement with the recent
estimate \cite{LT94}. These values correspond to a lower and upper limit of
the ONe nova rate of 2~\py\ and 18~\py\ , respectively.

\subsubsection{\label{s223} \na\ yield range}

The most recent upper-limit for the \na\ mass ejected by ONe novae 
is provided by Iyudin et al. \shortcite{Iyu95}, who derived 3~10$^{-8}$\Msol\ for any neon-type nova in the Galactic disk, at a 2$\sigma$ level with COMPTEL. Recent theoretical estimations \cite{JH98} provide typical \na\ yields of $\approx$2~10$^{-9}$~\Msol\ per ONe nova, well below such observational upper limit. However, the average yield of \na\ ejected per nova 
is not yet well known. It is sensitive to the composition of the 
underlying white dwarf and to the degree of mixing between the core and the accreted enveloppe (which constrain not only 
the ejected envelope mass but also the maximum temperature reached during the 
runaway) and also to the convection efficiency \cite{Star98}, \cite{JH98}. 
It is important to remind that in some cases theoretical models predict total 
ejected masses which are smaller than some observed ones (which in turn have 
often large determination uncertainties).
Moreover, some uncertainties of nuclear reaction cross-sections affect the 
computed \na\ yields: a net increase in the \na\ 
ejected mass by a typical ONe nova is obtained, with a very recent update of 
the nuclear reaction network affecting \na\ synthesis. The new 
\na\ yields for typical ONe novae are larger by factors ranging between 
$\sim$2 and $\sim$10, resulting in ejected masses as large as 10$^{-8}$~\Msol\ 
for a 1.25 \Msol\ ONe nova in the most favorable case of nuclear reaction
rates \cite{JCH99}.

\subsection{\label{s23} Results: maps of the 1.275 \MeV\ emission}

\begin{figure}
 \epsfig{file=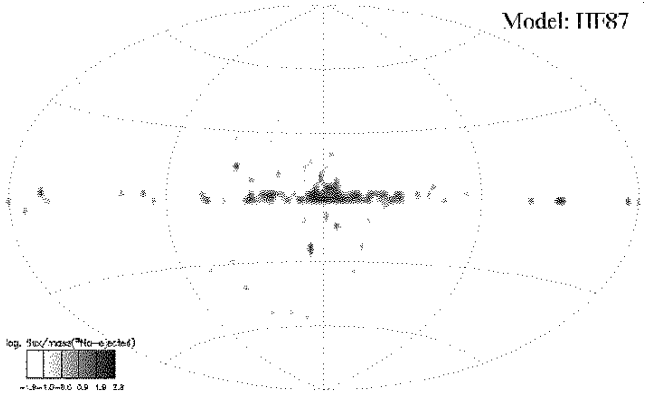,height=5.2cm,width=8.5cm} 
 \caption{ Example of modeled diffuse 
1.275 \MeV\ emission generated by a Monte Carlo simulation for the HF87 model. The fluxes have been normalized with the \na\ ejected mass (photons s$^{-1}$ 
cm$^{-2}$ \Msol$^{-1}$). The relative intensity is plotted with 3\deg\ by 
3\deg\ pixels in a map in galactic coordinates. For this example of Galaxy-test, the adopted ONe nova rate is $\approx$10 \py.}
 \label{fig:examHF87}
\end{figure}
\begin{figure}
 \epsfig{file=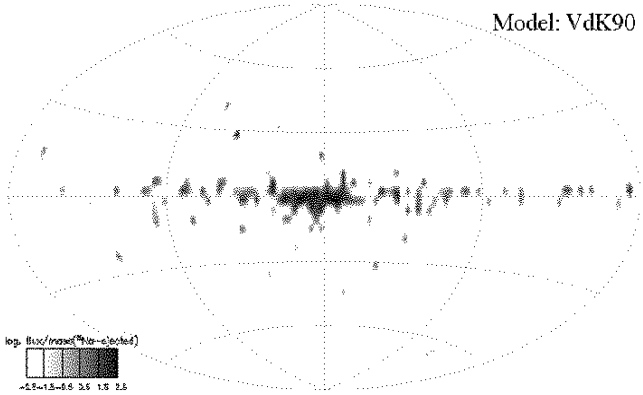,height=5.2cm,width=8.5cm} 
 \caption{Same as figure \ref{fig:examHF87} but for the VdK90 model.}
 \label{fig:examVdK90}
\end{figure}
\begin{figure}
 \epsfig{file=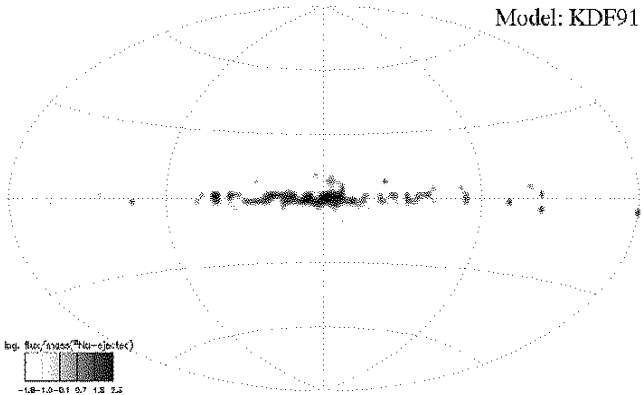,height=5.2cm,width=8.5cm} 
 \caption{ Same as figure \ref{fig:examHF87} but for the KDF91 model.}
 \label{fig:examKDF91}
\end{figure}
\begin{figure}
 \epsfig{file=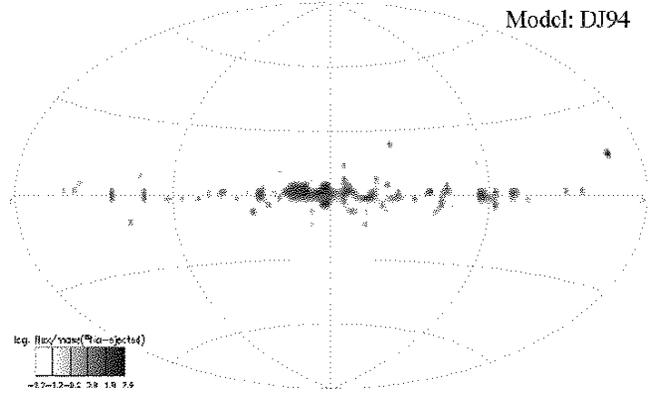,height=5.2cm,width=8.5cm} 
 \caption{Same as figure \ref{fig:examHF87} but for the DJ94 model.}
 \label{fig:examDJ94}
\end{figure}

Examples of Galaxy-tests for the distributions of novae in the Galaxy 
presented in section 2.2.1 are shown in Figures 
\ref{fig:examHF87}, \ref{fig:examVdK90}, \ref{fig:examKDF91} and 
\ref{fig:examDJ94}. For these simulations, the adopted Galactic nova rate is 
35 \py\ \cite{Sha97} and the proportion of ONe novae is 30\%. 
As already shown by HF87, the 1.275~\MeV\ intensity is irregularly distributed in the Galaxy. We have recovered the \na\ upper-limit mass found by HF87 with the sensitivity of HEAO3 - 4 10$^{-4}$ \funit\ at a 3$\sigma$ level, \cite{Mah82} - by simulating their frequency-spatial distribution of Galactic novae (rate 46 \py and spatial distribution as shown in Table \ref{tab:model})
and by integrating the flux between 330\deg\ and 30\deg\ longitude. 

\begin{figure}
 \epsfig{file=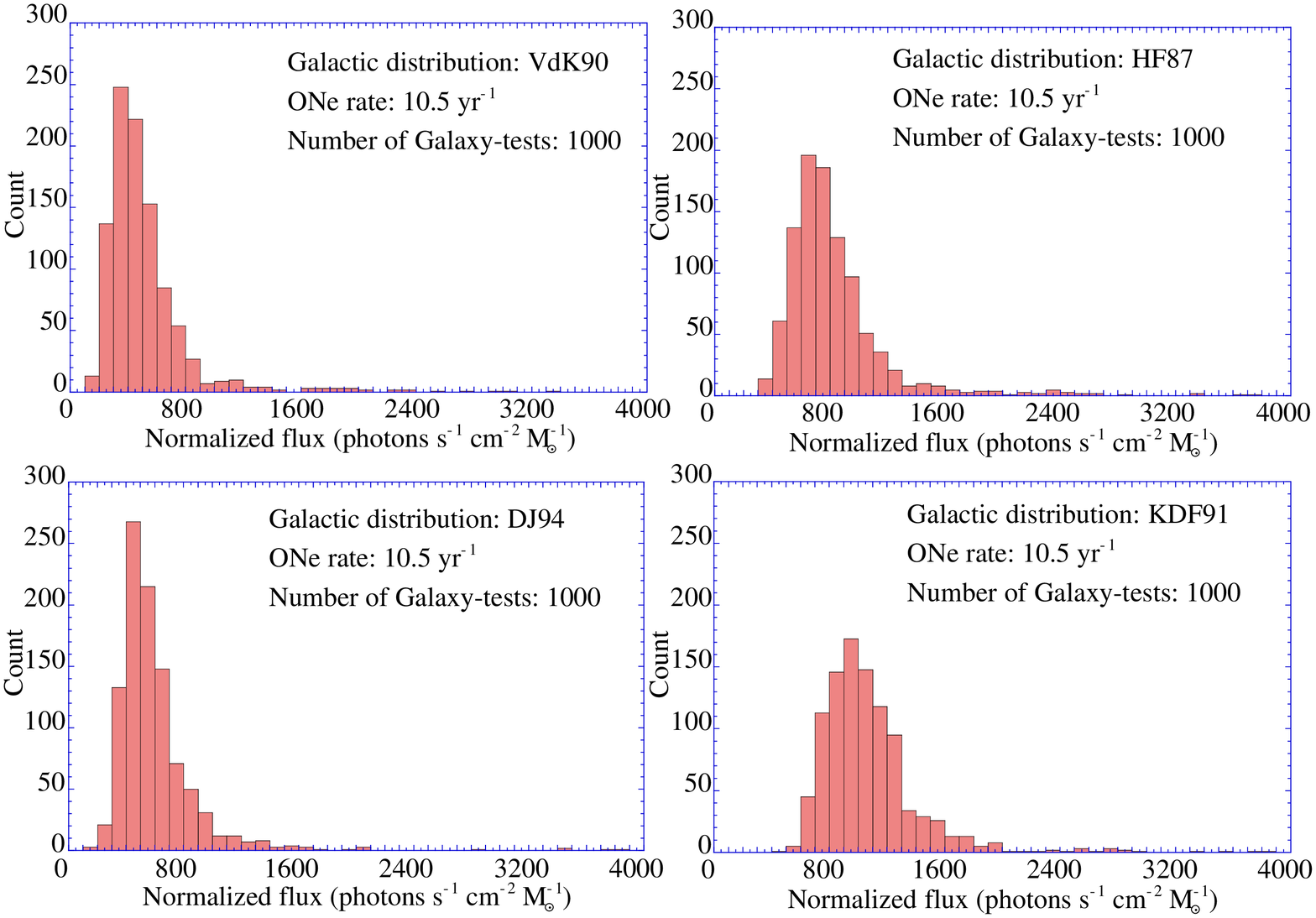,height=5.5cm,width=9.0cm} 
 \caption{Histogram of the normalized flux (flux per \na\ ejected mass per outburst) in the 12.5\deg\ around the GC obtained from 1000 simulated Galaxy-tests. The four models of spatial distribution are presented.}
 \label{fig:distflux}
\end{figure}

\begin{figure}
 \epsfig{file=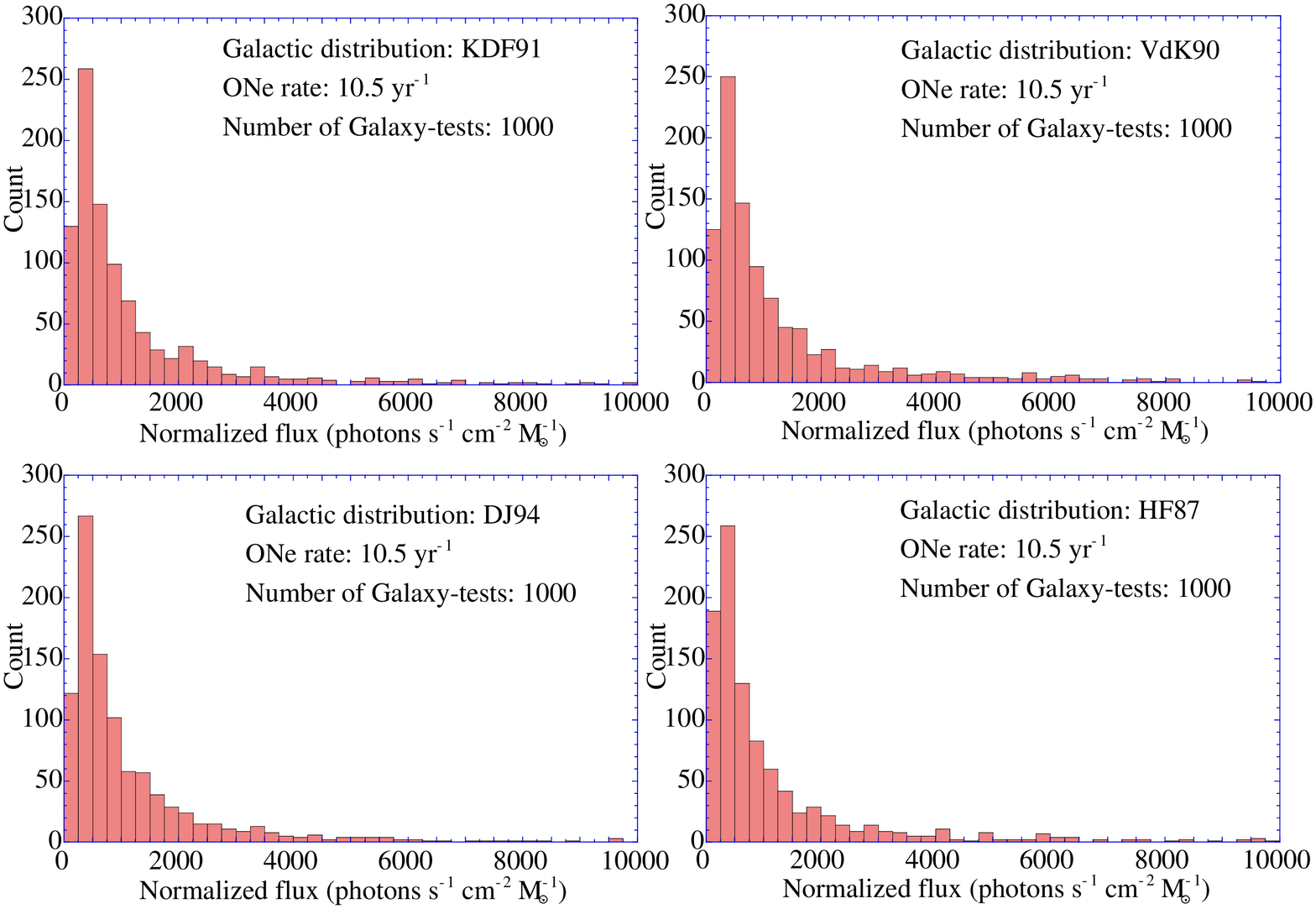,height=5.5cm,width=9.0cm} 
 \caption{Histogram of the normalized flux (flux per \na\ ejected mass per outburst) emitted by the brightest ONe novae of the Galaxy. The four models of spatial distribution are presented.}
 \label{fig:distbright}
\end{figure}

The statistical distributions of the 1.275 MeV fluxes from the 12.5\deg\ around
the GC and from the brightest nova of the Galaxy have been calculated using a 
large number of Galaxy-tests. The results are presented in figure 
\ref{fig:distflux} and \ref{fig:distbright} for an ONe nova rate of 
$\approx$10 yr$^{-1}$ and for the four models of spatial distribution. The 
KDF91 model provides the highest mean flux in the GC region, whereas 
the distribution of fluxes emitted by the brightest nova are 
similar in the four models, since the distributions of novae distances to the 
Sun are similar up to $\approx$7 kpc for all the models. 

Table \ref{tab:flux} shows the mean 1.275 \MeV\ flux from the GC entering in
the SPI field-of-view normalized to the \na\ mass ejected per year in the 
Galaxy. The uncertainties give an idea of the intensity fluctuation magnitude 
due to the limited number of effective emitting novae in the SPI field-of-view.
The normalized fluxes are directly related to the spatial characteristics of 
the chosen Galactic ONe novae distribution (see Table \ref{tab:model}). 
The total 1.275 \MeV\ normalized flux of the Galaxy is also presented in 
Table \ref{tab:flux}. The intensity in the GC region represents 15\%, 10\%, 
21\% and 11\% of the total flux in the Galaxy for the HF87, VdK90, KDF91 and 
DJ94 models, respectively.

\begin{table} 

\caption{Mean simulated 1.275~\MeV\ flux emitted by ONe novae in the 
12.5\deg\ around the GC region and in the whole Galaxy. 
The flux $F_N$ is normalized to the \na\ mass ejected per outburst and the rate of Galactic ONe novae (\funit \Msol$^{-1}$ yr). Values correspond to 
the four adopted models.}

\label{tab:flux}

\smallskip
\begin{tabular}{|l|cc|}
Models & $F_N(GC)$        & $F_N(Total)$    \\
HF87   & 78.0 $\pm$ 2.4   & 507 $\pm$ 85    \\
VdK90  & 56.8 $\pm$ 3.5   & 580 $\pm$ 287   \\
KDF91  & 105.1 $\pm$ 2.9  & 508 $\pm$ 62    \\
DJ94   & 56.4 $\pm$ 1.4   & 496 $\pm$ 119   \\ 

\end{tabular} 
 
\end{table}

We have investigated the variation of the normalized 1.275 \MeV\ intensity in 
the 12.5\deg\ around the GC as a function of the scaleheight and scalelength 
of the novae distributions in the case of a zero proportion of novae in the 
bulge (see below). For a 0.3 kpc scaleheight, the normalized intensity 
decreases from $\approx$89 to $\approx$31 \funit \Msol$^{-1}$ yr when the 
scalelength increases from 
2.5 to 6.0 kpc. Although the novae distances to the Sun are decreasing 
with increasing scalelength, the distribution around the GC is more diluted 
and the flux entering in the field-of-view is therefore lower.

\section{\label{s3} Analysis of the Galactic 1.275 \MeV\ emission by SPI}

\subsection{\label{s31} Types of detection}
 
\begin{figure}
 \epsfig{file=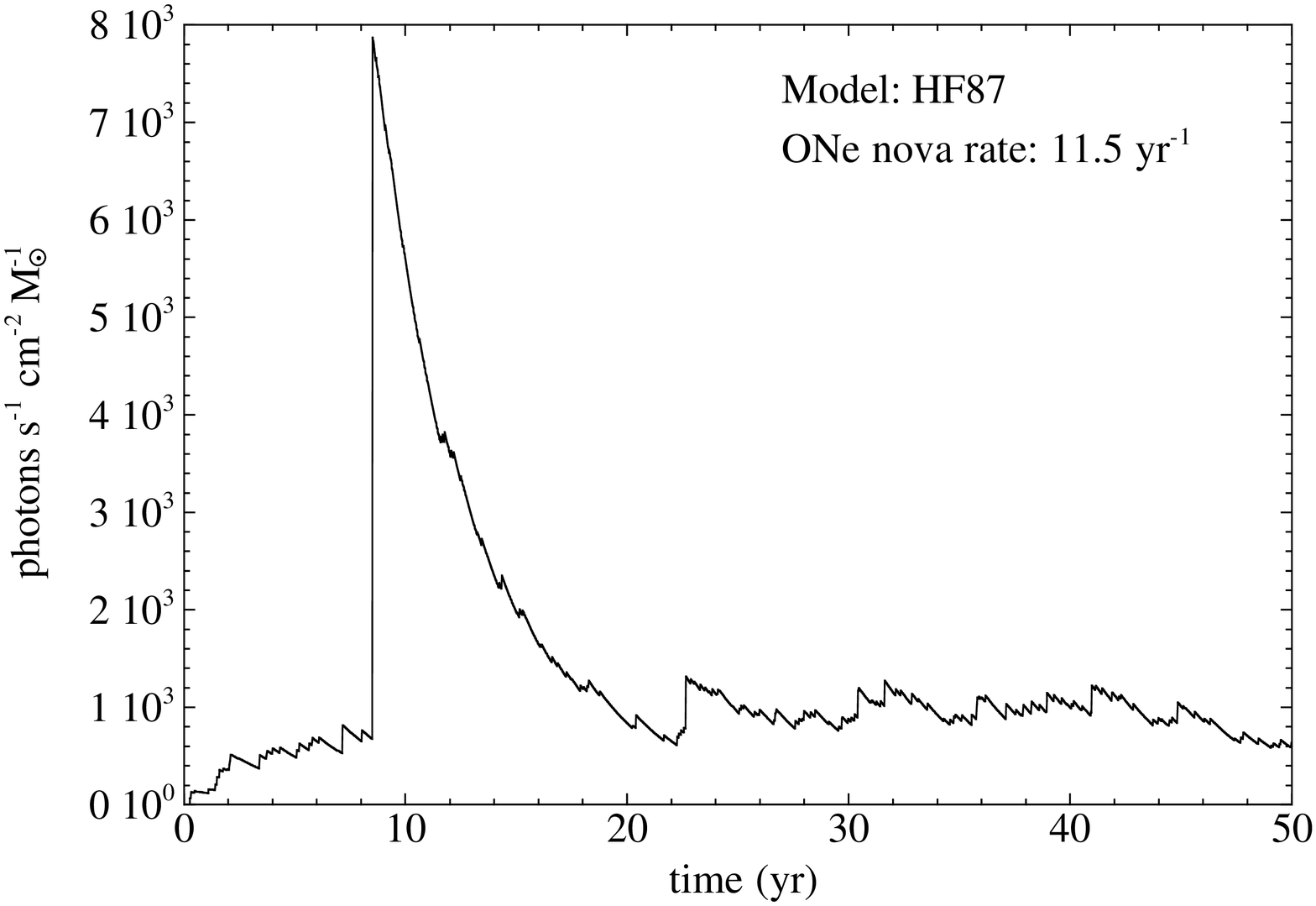,height=5.2cm,width=8.5cm} 
 \caption{ Example of the total 1.275~\MeV\ normalized flux in the 12.5\deg\ around the GC as a function of time. This results from a simulation with the HF87 model and an ONe nova rate of 11.5 yr$^{-1}$. The flux in the GC can be dominated by a single close nova.}
 \label{fig:timevar}
\end{figure}

\begin{figure}
 \epsfig{file=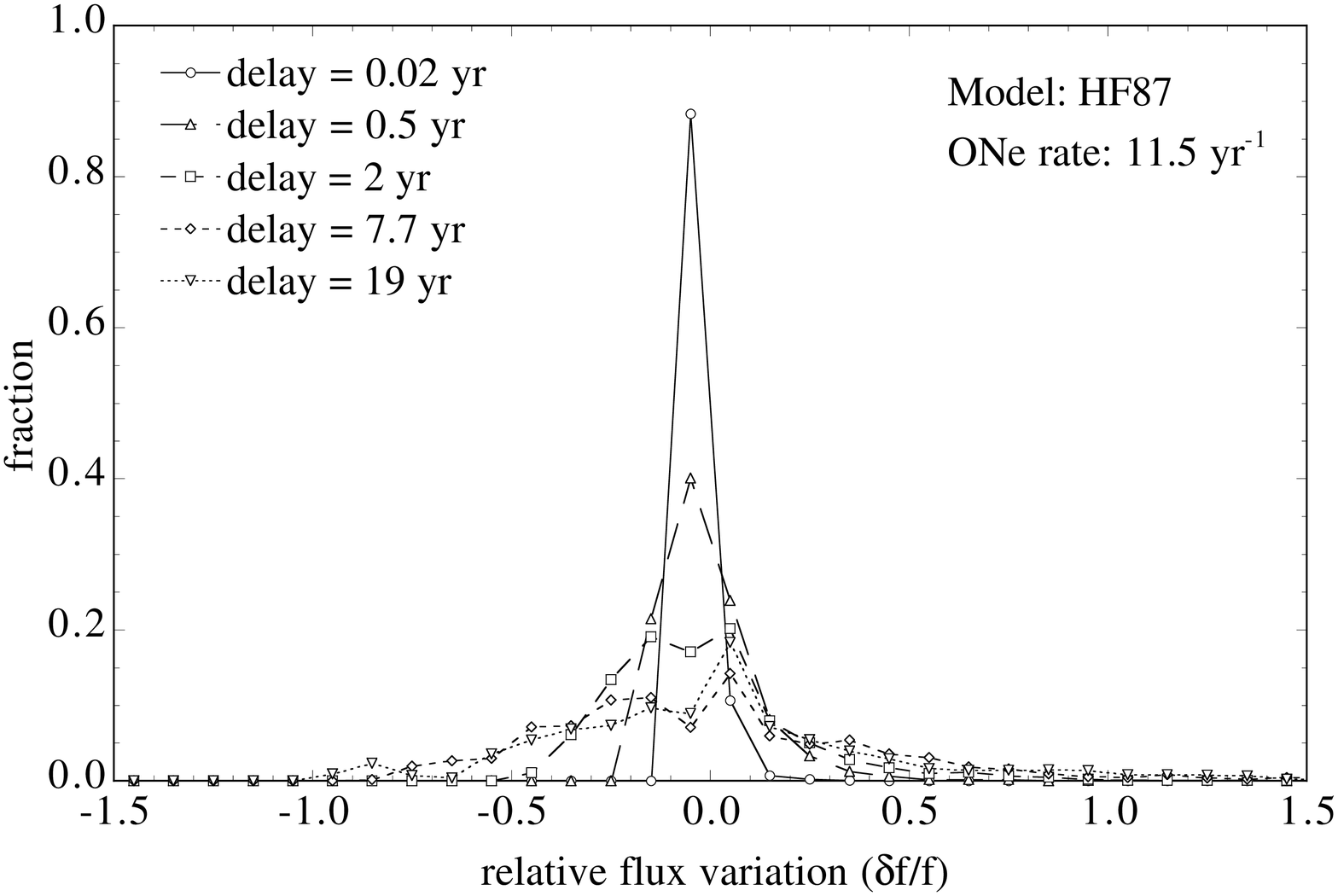,height=5.2cm,width=8.5cm} 
 \caption{Distributions of the relative variations of the 1.275~\MeV\ normalized flux in the 12.5\deg\ around the GC between two observations separated by delays of 0.02 yr, 0.5 yr, 2 yr, 7.7 yr and 19 yr. These distributions are obtained with the light curve presented in figure \ref{fig:timevar}.}
 \label{fig:dfdist}
\end{figure}

Each one of the above mentioned Galaxy-tests 
could be observed by a \gray\ spectrometer at a given date. This
is true if the observation time is short as compared with the occurrence period of ONe novae and with the time variation of the \na\ emissivity. These
conditions are satisfied since the common observation time of the future 
\gray\ spectrometer SPI will be $\approx$11~days, that is shorter than both the 
average ONe novae occurrence period ($\approx$45~days) in the whole Galaxy, 
and the decay period of \na\ (3.75~yr, implying 1.275~\MeV\ flux 
reduction of only 0.8\%\ in 11~days). However, for an observation time longer 
than few times the ONe novae occurrence period, the time variation of the
1.275~\MeV\ Galactic emissivity has to be accounted for. 
Figure \ref{fig:timevar} shows an example of the total 1.275~\MeV\ flux 
variations in the 12.5\deg\ around the GC. It results from a Monte Carlo 
simulation as described in the previous section. The flux has been normalized 
to the \na\ ejected mass per outburst. In figure \ref{fig:timevar}, the 
flux in the GC is dominated by a single nova (which contributes to 91\% of the 
total flux) at 8.6 yr. After 20 yr, the flux shows 
fluctuations with a maximum amplitude of $\pm$40\% of the mean flux. 
Figure \ref{fig:dfdist} shows the statistical distributions of the relative 
variations between two epochs of the total 1.275~\MeV\ flux in the 12.5\deg\ 
around the GC. They have been calculated using the flux variation of 
figure \ref{fig:timevar}. Several delays between the two epochs are presented. 
For two observations 6 months apart, there are probabilities of 15\% and 21\% 
to get a flux variation of more than 10\% and less than -10\% respectively. 
Even if the flux variations are not large, the weak additional spots that 
could appear in the field-of-view will modify the distribution of the 
pixel-intensity. This could modify significantly the analysis in the case of 
the detection of excesses in the spatial distribution (see section 3.1.2).



As stated above (section 2.1), the number of novae and the distribution 
of the intensity have been calculated with 3\deg\ by 3\deg\ pixels. Therefore, 
several novae can be located in the same pixel (this happens mostly in
the direction of the GC region). However, the flux in those pixels can
be dominated by only one nova. In this case, the measured intensity
can not be considered as a cumulative effect. Therefore, we state for a
seldom-nova criterium when a single nova contributes to more than
95\%\ of the total flux in a pixel.

Jean \shortcite{Jea96} and Jean et al. \shortcite{Jea97} have estimated the narrow line sensitivity of SPI for an on-axis point source by computing instrumental background and detection 
efficiency with Monte-Carlo simulations including detailed physics. However,
G\'omez-Gomar et al. \shortcite{Gom98} have shown that the 1.275 \MeV\ line should have a typical width of 20~\keV\ (FWHM). With these conditions, the sensitivity is 2.2 10$^{-5}$ \funit\ at a 3$\sigma$ level and for 10$^6$ s ($\approx$11 days) of observation time, instead of 6 10$^{-6}$ \funit. In the present case, the Galactic ONe novae emission is diffuse and the spectrometer will also provide a distribution of intensity in pixels.  

Therefore, three types of detection have been investigated for the analysis of 
any Galaxy-test. 

\subsubsection{\label{s311} Detection of the cumulative emission at 
1.275 \MeV\ (case 1)} 

The total 1.275 \MeV\ flux impinging on the SPI detectors will generate counts
that will be added to the instrumental background ones. It will be possible
to estimate these source counts by substracting the instrumental background
component from the overall detector rates without using imaging
deconvolution. Contrary to the imaging method, where the background rate is
estimated using detectors that are not illuminated by a source (i.e. half
of the total number of detectors - which are below opaque mask elements),
the instrumental background is estimated using either observations pointing
out of the galactic plane - i.e. the on/off method used in balloon-borne 
spectrometers - or background modelling. The latter can consist of estimating 
the instrumental background under the line of interest either by using data 
obtained during the observation (rate around the studied line, other 
background feature rates), or by fitting the background temporal variations 
before and after the observations, or both of the two methods.

For this study, we simply consider the measurement of the cumulative emission 
with the on/off method. The sensitivity is estimated using the calculation of 
Jean \shortcite{Jea96} (see also SPI - Science Performance Report \shortcite{SPIrep}) assuming that the time spent to measure instrumental background is equal to the GC observation time and that there is no more modulation of the source photons by the coded aperture. In this case the exposure is on average equal to the point source exposure since half of the flux irradiates the detection plane (the other half being hidden by the 
coded mask elements). Since the counts recorded over all the detectors are 
taken into account, the background is increased by a factor $\approx$2 with 
respect to the imaging case. Therefore, since the sensitivity increases with 
the square root of the background, the sensitivity of such a mode of detection 
is roughly the sensitivity for an on-axis point source (in the imaging case) 
multiplied by the square-root of 2 (i.e. 3.1 10$^{-5}$ \funit\ at a 3$\sigma$ 
level and for 10$^6$ s of observation time). In this method, the single 
information that will be available is the number of detected 1.275 \MeV\ 
photons coming from the region in the field-of-view of SPI. Although such a detection will give the proof of \na\ decay in the Galaxy, it will not provide any information on its spatial distribution in the observed region. However, galactic \na\ yield can be estimated provided that assumptions on the ONe distribution in the Galaxy are made.

\subsubsection{\label{s312} Detection of excesses in the spatial distribution 
of the intensity at 1.275 \MeV\ (case 2)} 

The total flux in the SPI field-of-view is distributed in several pixels. Some 
of the pixels can contain an intensity that leads to a significant signal. In 
this case, we obtain an additional information that is the location of the 
emission in the Galaxy. However, it is necessary to account for the reduction 
of the significance of the detection with the number of pixels involved in the 
total flux. A rough estimation of the significance (number of~$\sigma$) of 
such a detection is derived by computing the probability that the observed 
intensity distribution in the field-of-view is due to background fluctuation. 
The lower the probability, the higher the significance. The number of sigma 
for each pixel is calculated by comparing the flux with the sensitivity for 
a point source \cite{Jea97}. Afterwards, this number of sigma is converted in 
probabilities. The combination of the probabilities of the overall pixels 
allows us to estimate the probability that the observed distribution is not due 
to background fluctuation. Equation \ref{eq:pr} shows an estimation of this 
probability $P$ that is available when the significances in pixels are high 
enough. $N$ is the total number of pixels in the SPI field-of-view, $k$ is the 
number of pixels above a given significance threshold (2.5$\sigma$ in this 
work) and $P_i$ is the probability derived from the significance in the pixel 
$i$.

\begin{equation}
 P \; = \; \frac{N!}{(N-k)!} \; \prod^k_{i=1} P_i
\label{eq:pr} 
\end{equation}

The distribution of the emission from the GC is considered to be detected when 
the calculated probability leads to a significance that is above 3$\sigma$ for 
an observation time of 10$^6$ seconds ($\approx$11 days). A single nova close 
to the Earth can be the major source in the SPI GC field-of-view. Account for 
such an event makes a bias on the statistics of the cumulative 1.275~\MeV\ flux. Therefore, to avoid such effect, the effective single nova pixel is removed from the analysis if its flux is more than 95\%\ of the central 25\deg\ total flux.

\subsubsection{\label{s313} Detection of the brightest nova of the Galaxy in 
1.275 \MeV\ (case 3)} 

The flux from the brightest nova that can be observed from the Earth is also 
analyzed and considered to be detected if its value is above the 
sensitivity of the spectrometer SPI, for an observation time of $\approx$11 
days. The observation time needed to detect with SPI (at a 3$\sigma$ level) the brightest nova of the Galaxy-test has also been estimated. Equation~(\ref{eq:tf}) allows us to calculate the duration of 
the observation ($T_{3\sigma}$ in days) to get a 3$\sigma$ detection of 
the flux $f$ (\funit) in a single pixel.
 
\begin{equation}
 T_{3\sigma} \; = \; 2.42 \: 10^{-5} \; f^{-1} \; + \; 5.28 \: 10^{-9} \;
 f^{-2} 
\label{eq:tf} 
\end{equation}

We can recognize the well known variation of the sensitivity as a
function of the square root of the observation time in the second term
of the equation, that is dominant when the signal-to-noise ratio is
low (i.e. low fluxes). The first term accounts for the statistical
fluctuations induced by the counts from sources when the
signal-to-noise ratio is not low anymore (i.e. for high fluxes).

\subsection{\label{s32} Probability to detect the 1.275 \MeV\ line with SPI}
 
For a large number of Galaxy-tests, the probability to detect any type of 
emission can be estimated by calculating the fraction of time that the 
simulated 1.275~\MeV\ flux is above the
upper-limit of a given experiment (HEAO3 or SPI). For the particular case of 
the future spectrometer of INTEGRAL, the average times of observation to
obtain detections of the three types mentioned above
have been derived, as well as the average number of novae that contribute to 
the diffuse emission of the GC region.
Each Galaxy-test has been analyzed in order to check whether SPI would detect: 
the cumulative emission at 1.275~\MeV\ in the 12.5\deg\ around the GC 
region (case 1), an excess in the distribution of the emission (case 2) and 
the brightest nova of the Galaxy (case 3), for a given average \na\ 
yield per ONe nova outburst. The time needed for a 3$\sigma$ level detection 
is also computed. Two hundred Galaxy-tests 
have been simulated for each model and for a given \na\ yield in order to 
calculate the number of Galaxy-tests that lead to a detection by SPI with 
an observation time of 10$^6$ s. The number obtained, divided by the total 
number of Galaxy-tests, gives an estimation of the probability to detect 
the 1.275~\MeV\ line from ONe novae with SPI.

\begin{figure*}
 \epsfig{file=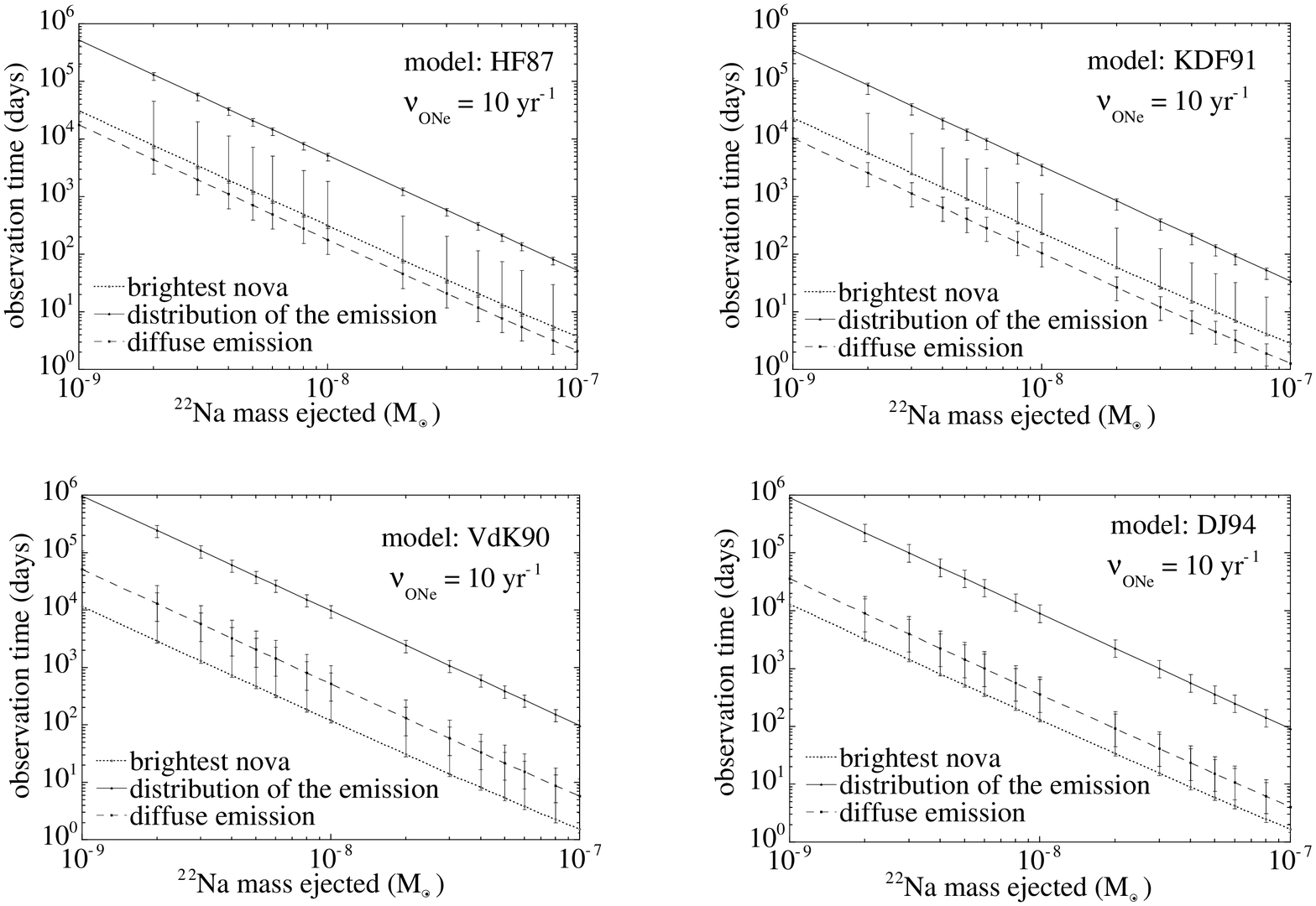,height=10.0cm,width=16.4cm} 
 \caption{Mean observation time needed to detect the cumulative (diffuse) emission in the GC (case 1), the distribution of the emission (case 2) and the brightest nova (case 3) at 1.275~\MeV\ with the future spectrometer SPI, as a function of the mean \na\ yield per nova. The error bars have a width of 1$\sigma$. The adopted Galactic ONe nova rate is 10~\py. Each point results from the processing of 200 Galaxy-tests. Each panel corresponds to a different spatial distribution of novae (see text for details).}
 \label{fig:timeObs}

 \vspace{0.1cm}

 \epsfig{file=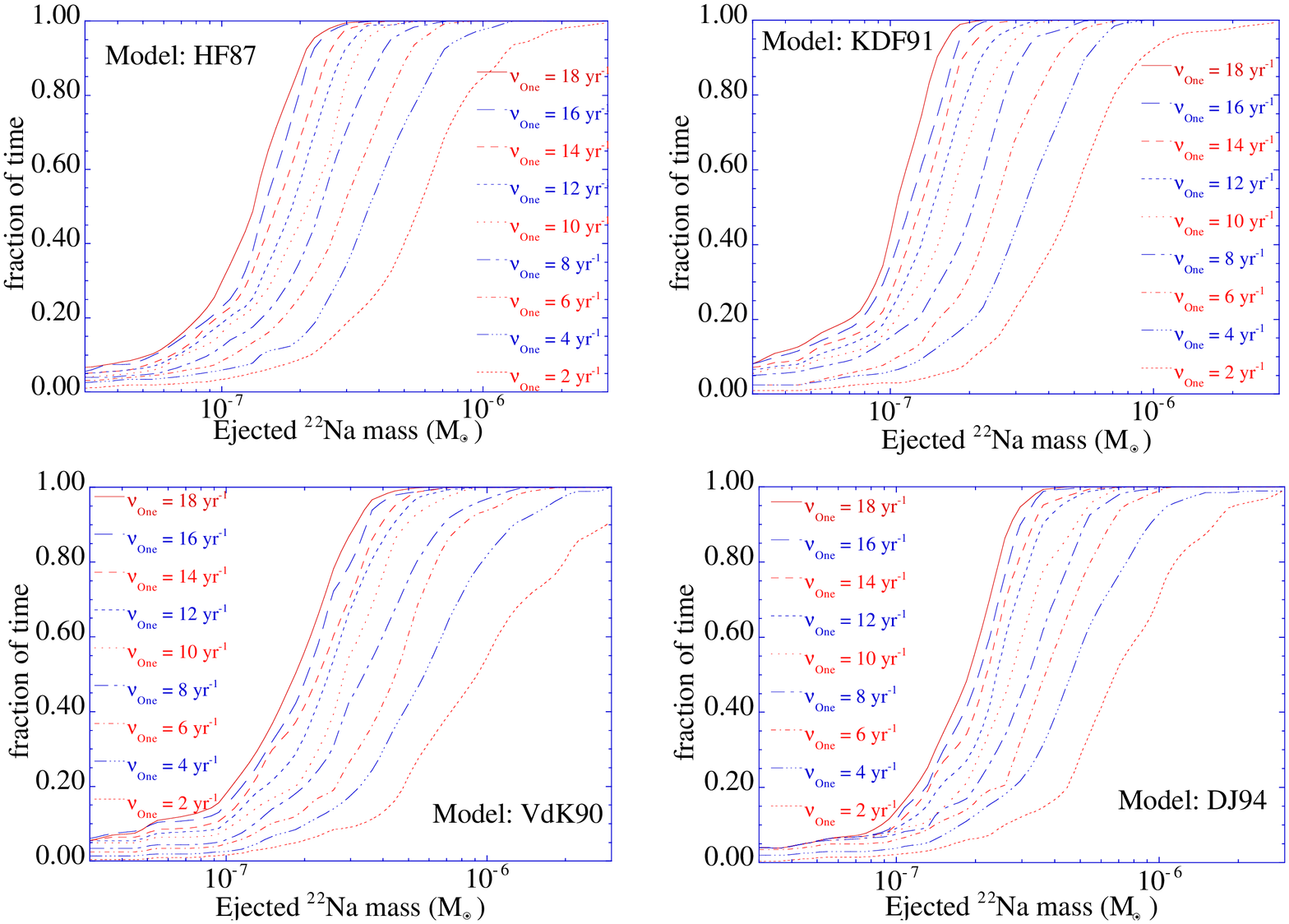,height=10.0cm,width=16.4cm} 
 \caption{Fraction of time during which the distribution of the diffuse emission, in a sphere of 25\deg\ diameter around the GC would be detected by SPI (case 2), as a function of mean \na\ yields per nova, calculated for ONe nova rates ranging from 2 to 18 \py and for the four adopted Galaxy models. Each point results from the processing of 200 Galaxy-tests.}
 \label{fig:fracTime}

\end{figure*}

Figure \ref{fig:timeObs} shows the mean observation times for the three types
of detection with SPI, as a function of the
average \na\ mass ejected per nova. The adopted ONe nova rate is
10~\py and the spatial distributions of novae are those described in section
2.2.1. For low values of \na\ yield, the observation times are very long
and do not make sense since the limited duration of the INTEGRAL mission
does not allow for observation times larger than 2 years.
Moreover, as the Galactic 1.275~\MeV\ intensity and its distribution would
change due to the decay of \na\ and to the explosion of additional novae, 
the analysis of the intensity distribution (case 2) and of the brightest nova flux (case 3) are not valid anymore. However, other analyses (cases 1) can be possible with exposures in the limit of the allocated observation time.

Figure \ref{fig:fracTime} shows the fraction of time that 
SPI would detect, in $\approx$11 days, an excess in the distribution of the 
emission from the GC region (case 2) as a function of 
average \na\ masses ejected per nova, for Galactic ONe nova rates ranging 
from 2 to 18 \py (i.e., an estimation of the probability to detect this 
emission).
In this analysis, it is not possible to normalize the
\na\ mass ejected by the ONe nova rate, because when the 
frequency of novae increases, the ($l, b$) distribution of intensity changes, 
resulting in an estimation of the significance that implies more pixels 
and not only more flux. Therefore, the sensitivity to an excess in the 
1.275~\MeV\ intensity distribution is not directly proportional to the amount 
of \na\ ejected per year.

\begin{figure*}
 \epsfig{file=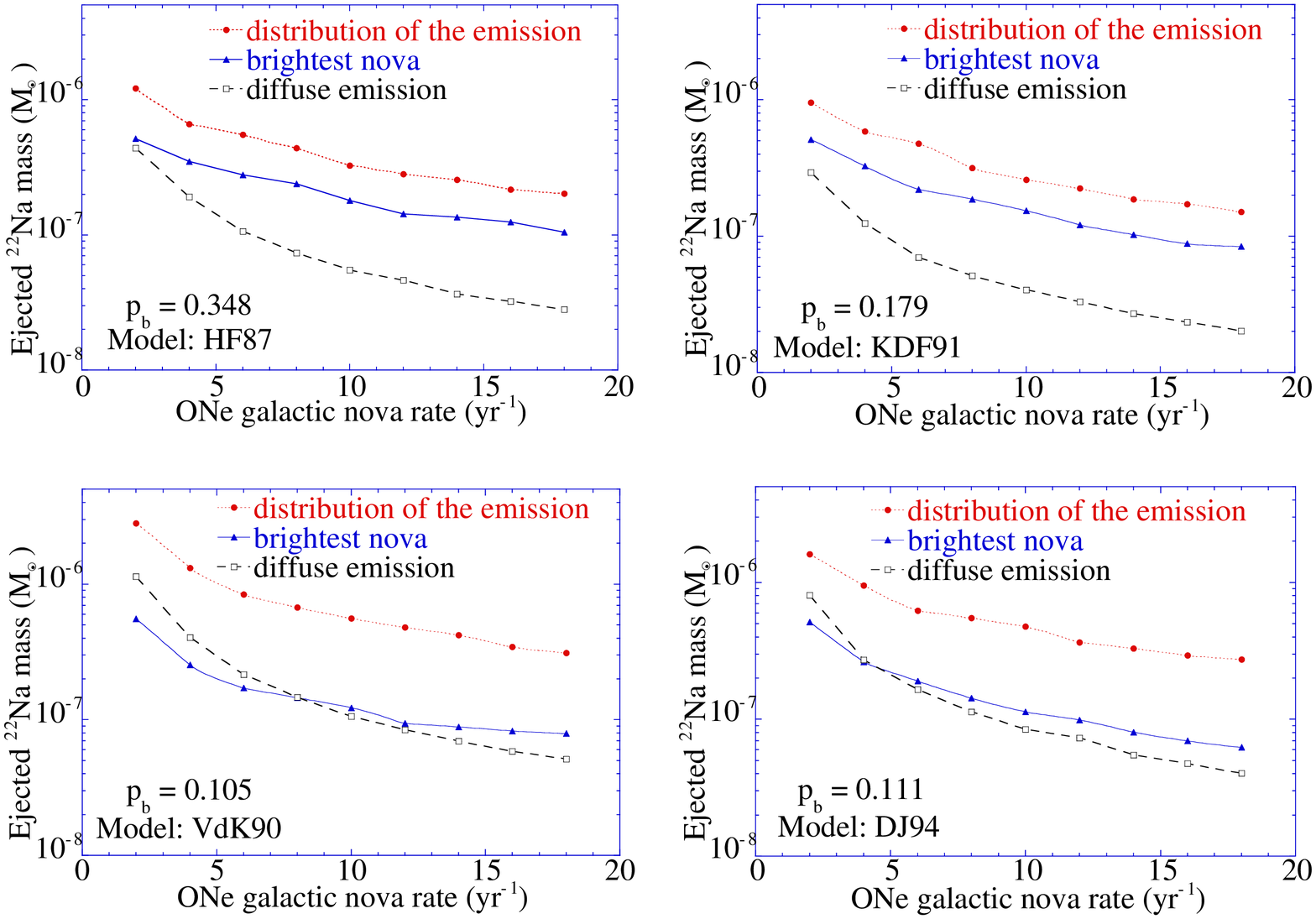,height=10.0cm,width=16.4cm} 
 \caption{Lower-limits of the \na\ mass ejected per nova that can lead to 90\%\ of probability that the 1.275~\MeV\ emission, from the novae in the GC region (total flux -case 1- and distribution -case 2) and from the brightest nova of the Galaxy (case 3), will be detected with SPI at a 3$\sigma$ level in 10$^6$ s of observation, as a function of the ONe nova rate.}
 \label{fig:LowerLimit}
\end{figure*}

Our results allow for the determination of the \na\ yield lower-limit that can 
be detected by the spectrometer of INTEGRAL. We state that this lower-limit 
of \na\ ejected mass corresponds to a probability of detection of 90\%. 
The lower-limit depends also on the nova rate, since more ONe novae lead to 
higher cumulative GC region flux, on the Galaxy model and on the type of detection. Taking the \na\ mass values at 90\%\ of 
probability in Figure \ref{fig:fracTime} we can calculate the lower-limit 
as a function of the Galactic ONe nova rate for case 2. Similar calculations 
have been performed for the detection of the cumulative emission in 
the SPI field-of-view (case 1) and of the brightest nova of the 
Galaxy (case 3). The results are displayed in Figures \ref{fig:LowerLimit} 
for the four adopted spatial distributions, as a function of the nova rate. 
As it has already been shown in Figure 
\ref{fig:timeObs}, it is clear that the future spectrometer SPI will be 
more sensitive to the brightest nova of the Galaxy and the total diffuse 
flux than to a detection of an excess in the distribution of the emission.                               
The \na\ lower-limit masses needed to detect the brightest nova (case 3) are 
similar for the four models, since the distributions of novae distances 
to the Sun associated to these models are similar up to $\approx$7 kpc from 
the Sun. Concerning the \na\ lower-limit masses for the cumulative emission 
(case 1), it can be seen in Figure \ref{fig:LowerLimit} that models 
KDF91 and HF87 provide lower values than VdK90 and DJ94 models.
The radial scalelengths for the former are 3.0~kpc and 3.5~kpc, respectively, whereas the models VdK90 and DJ94 have higher radial scalelengths, 5~kpc in both cases (see Table \ref{tab:model}). Since novae are closer to the GC in the KDF91 and HF87 models, their cumulative 1.275~\MeV\ flux is higher and more compact (see Figures \ref{fig:examHF87}, \ref{fig:examVdK90}, \ref{fig:examKDF91} and \ref{fig:examDJ94}). 
For a given ONe nova rate, the \na\ lower-limit mass decreases with decreasing 
scaleheight of the Galactic disk: the lower the z$_h$ (i.e., models HF87 
and KDF91), the higher the cumulative 1.275~\MeV\ flux at the GC (see Table 
\ref{tab:flux}) and the smaller the \na\ lower-limit mass (see Figure 
\ref{fig:LowerLimit}). However, this variation is not significant: for 
instance, when 
the scaleheight increases from 0.1~kpc to 0.45~kpc the \na\ lower-limit 
mass increases from $\approx$7 10$^{-7}$ \Msol\ to 9 10$^{-7}$ \Msol, if 
we use the DJ94 model with variable scaleheight. Moreover, the limited angular 
resolution of SPI 
constrains to a lower-limit the observable disk scaleheight. Indeed, 
in the case of an exponential disk model (without bulge contribution), 
simulations show that 
the estimated \na\ lower-limit mass does not change for z$_h$ values lower 
than $\approx$0.15~kpc and $\approx$0.25~kpc, which correspond roughly to a 
scale-angle of 3\deg\ for a Galactic disk radial scalelength of 5~kpc and 
3~kpc, respectively.

\begin{figure*}
 \epsfig{file=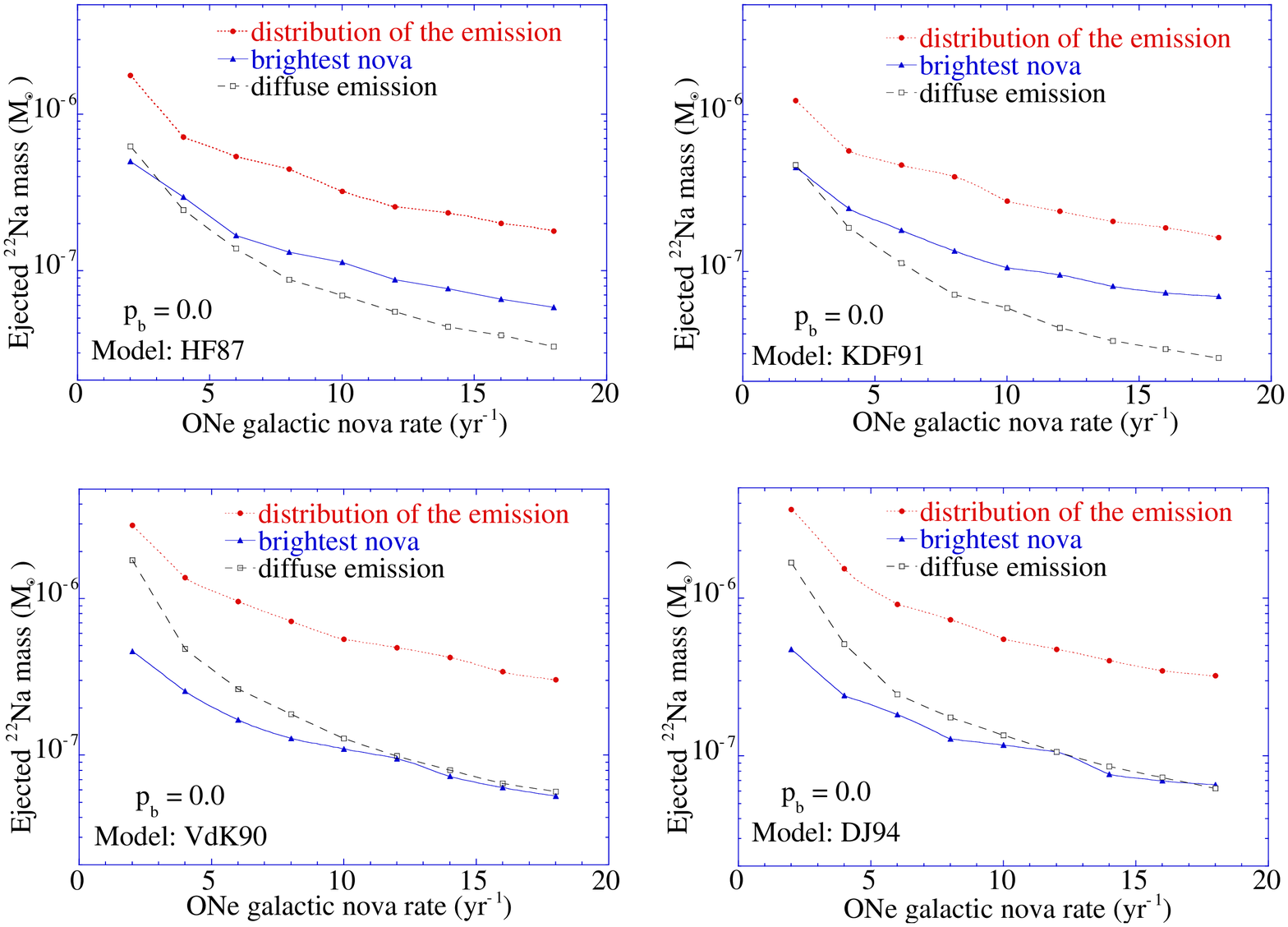,height=10.0cm,width=16.4cm} 
 \caption{Same as figure \ref{fig:LowerLimit} but without ONe novae occurring in the bulge.}
 \label{fig:LowerLimPb0}
\end{figure*}

It is not clear whether the Galactic bulge should be a source of 
1.275~\MeV\ photons, since it has been proposed that it is 10$^{10}$ years 
old and white dwarfs responsible of ONe novae should be much younger (the 
zero age mass of the white dwarf progenitor being probably larger than
$\sim$10 \Msol). However, the ages 
of novae do not only depend on the main sequence lifetime of the 
progenitor but also on the time necessary for the onset of the Roche lobe 
overflow in the binary system and the time between the onset of accretion 
and the outburst phase. Taking into account the eventuality that there is a 
lack of ONe novae in the bulge (see the recent paper by Della \& Livio \shortcite{DVL98}), 
we have estimated the \na\ mass lower-limits 
in the case of ONe novae only distributed in the Galactic disk. Figure 
\ref{fig:LowerLimPb0} 
shows the results of the simulations in this case for the four models adopted:
the \na\ mass lower-limits that SPI can detect are in general lower,
since novae are concentrated in the disk and, therefore, they are closer to
the Earth.

\begin{figure}
 \epsfig{file=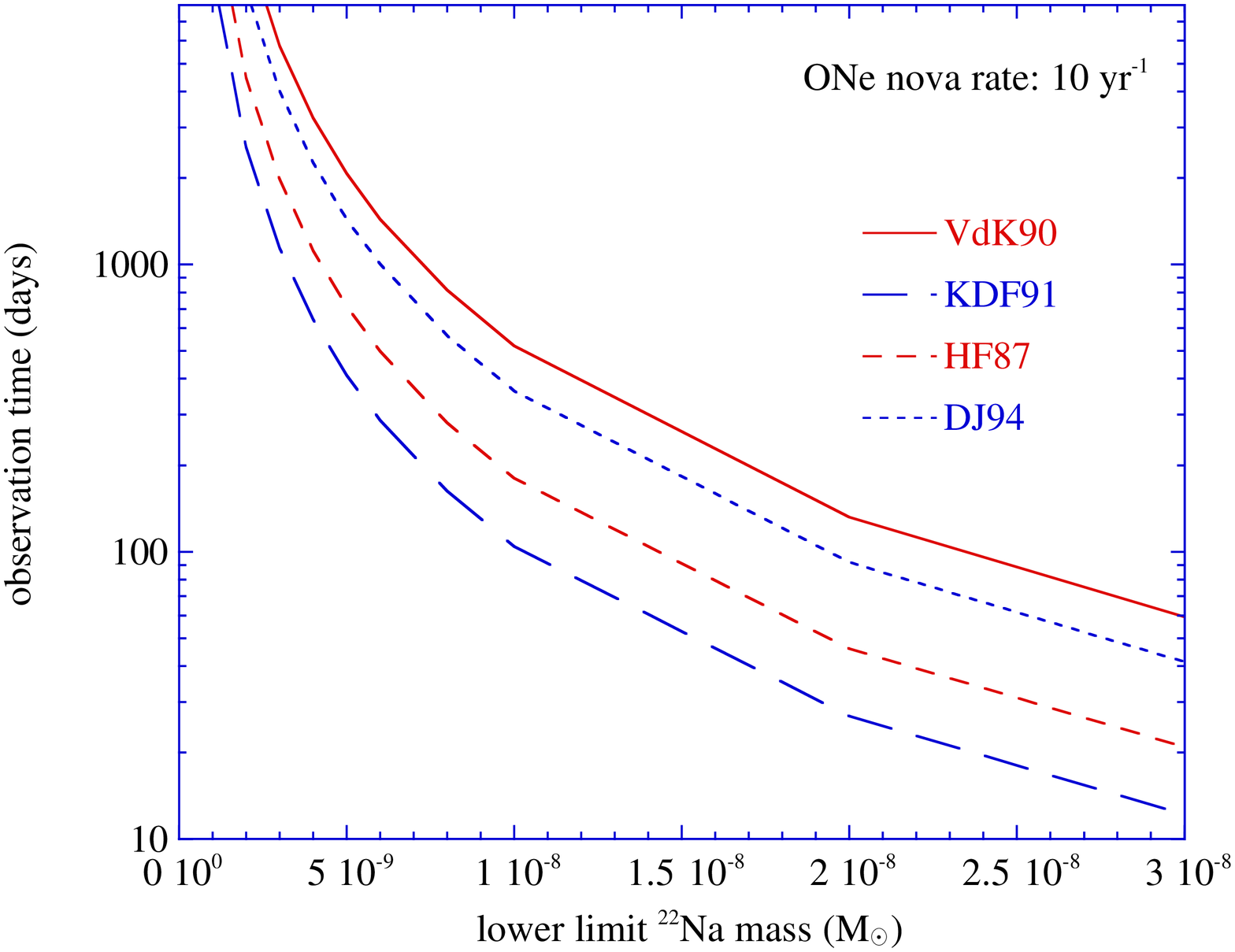,height=4.9cm,width=8.0cm} 
 \caption{SPI observation time necessary to have 90\%\ of chance to detect the cumulative 1.275 \MeV\ emission of the GC (case 1) as a function of the mean \na\ mass ejected per nova and for several Galactic distributions.}
 \label{fig:Tobs-vs-LowLim}
\end{figure}

As a final result, extracted from those presented in Figure \ref{fig:timeObs},  
Figure \ref{fig:Tobs-vs-LowLim} shows the observation time needed to have 
90\%\ of chance to detect with SPI the 1.275 \MeV\ line emitted by the 
accumulated \na\ from the GC region (case 1) as a function of the mean \na\ 
mass ejected 
per ONe nova. These results are displayed for the four distributions of novae 
in the Galaxy. For the HF87 and KDF91 spatial distributions and our
estimations of \na\ yields, observation times compatible with those planned 
for the deep survey of the central Galaxy are obtained. Therefore, cumulative 
emission in the GC may be detected with SPI.

\section{\label{s4}Discussion and Conclusions} 

\begin{figure}
 \epsfig{file=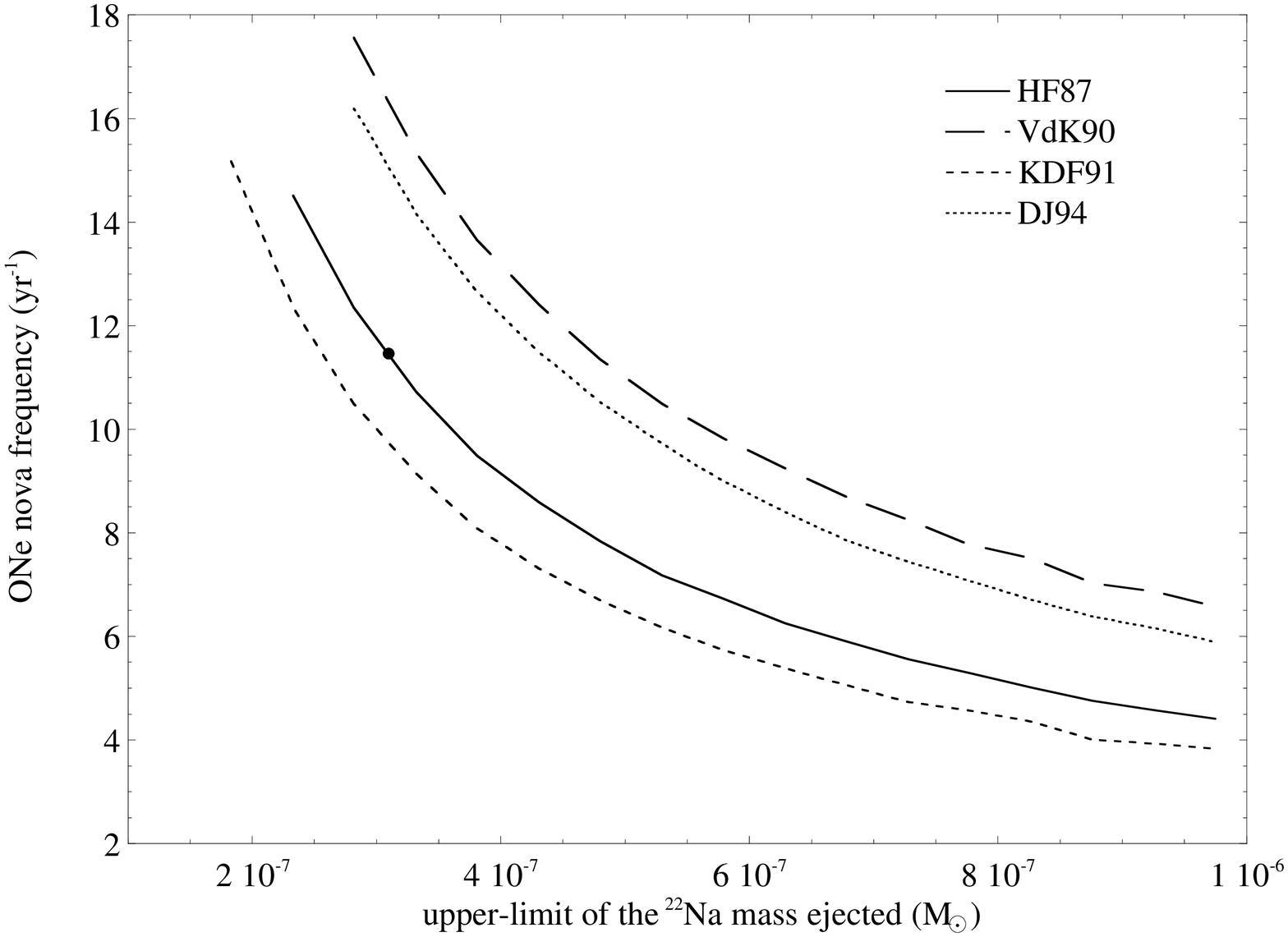,height=4.9cm,width=8.0cm} 
 \caption{Galactic ONe nova rate needed to have 90\%\ of chance to detect the cumulative 1.275 \MeV\ emission from the GC region (between 330\deg\ and 30\deg) with HEAO3 as a function of the mean \na\ mass ejected per nova and for several Galactic distributions. The dot shows the estimation of HF87.}
 \label{fig:heao-UppLim}
\end{figure}

It is important to remind that the total Galactic flux at 1.275~\MeV\ depends not only on the amount of \na\ ejected per outburst and the ONe novae rate but also on their distribution in the Galaxy. The ejected \na\ mass upper-limit derived by HF87 using HEAO3 measurement was estimated for only one distribution and an ONe nova rate of 11.5 yr$^{-1}$. Figure \ref{fig:heao-UppLim} shows the \na\ mass upper-limits computed with the 1.275~\MeV\ upper-limit flux of HEAO3 considering other possible frequency-spatial distributions of novae in the Galaxy. It shows that the \na\ mass upper-limit lies between 2 and 10 10$^{-7}$ \Msol\ per nova.

The spectrometer of INTEGRAL will be more sensitive for the detection of the 
brightest nova of the Galaxy and of the cumulative diffuse emission of ONe novae in the GC region by an ``on-off'' type observation than for the determination 
of their intensity distribution. But it is not clear whether the brightest 
novae will be first detected in the visible due to the interstellar extinction.
Nevertheless, long exposures in the Galactic plane region may allow for their 
detection with SPI. 

According to Jos\'e, Coc \& Hernanz \shortcite{JCH99}, the \na\ average yield could be between 3.~10$^{-9}$ and 1.2~10$^{-8}$ \Msol\ per nova. Therefore, there are few chances to detect the Galactic diffuse 1.275~\MeV\ emission with SPI with only $\approx$10 days of observation. However, 80~days of observation of the GC region could already give constraints to the mean \na\ yield in ONe novae as a function of their frequency-spatial distribution in the Galaxy (see Figure \ref{fig:Tobs-vs-LowLim}). 

It is worth mentioning that the adopted width of the 1.275~\MeV\ line
(20~\keV) is quite large, corresponding to the fastest ONe novae, which have 
the largest ejection velocities (and also the largest ejected \na\ masses). 
As the correlation between \na\ ejected mass and ejection velocity is not 
well established (neither theoretically nor observationally), and the braking 
of the ejected mass during the lifetime of \na\ is in principle small but not 
null, we are in the 
worst situation when we adopt such broad lines (corresponding to the largest 
velocities and without any braking effect). In the other extreme case of 
narrow lines (less realistic but interesting to make a comparison), the
sensitivity of SPI is 6 10$^{-6}$ \funit\ at a 3$\sigma$ level and for 
10$^6$ s ($\approx$11 days) of observation time, for a point source,  
which is $\sim$3.7 times better 
than the one for our broad lines (see section \ref{s31}) and allows for the
detection of lower fluxes by the same factor. Consequently, a positive 
detection of the 1.275~\MeV\ line by SPI would be achieved either with a 
smaller \na\ mass or with a smaller observation time.

Gamma-ray observation of novae would provide information not only on their eruption mechanisms and the nucleosynthesis processes involved in their explosion but also on their distribution and their rate in the Galaxy (e.g. proportion in the bulge, scaleheight in the Galactic disk) since the problem of the interstellar extinction does not appear at this energy range.

%
%
\section*{Acknowledgments}

We wish to thank G. Skinner for his helpful comments.

Research partially supported by the training and Mobility Researchers 
Programme, `Access to supercomputing facilities for european researchers', 
established between the European Community and CESCA-CEPBA, 
under contract ERBFMGECT050062, and with the 
research projects ESP98-1348 (CICYT-PNIE), PB97-0983-C03-02 and 
PB97-0983-C03-03 (DGICYT), GRQ94-8001 (CIRIT).
%
%
%
%

\label{lastpage}

\end{document}